\documentclass{svproc}
\usepackage{url}

\usepackage{siunitx}
\usepackage{color}
\usepackage[table]{xcolor} 
\usepackage{subcaption}
\usepackage{tabularx}
\usepackage{booktabs}
\usepackage{floatrow}
\usepackage{float}
\usepackage{multirow, array}

\newcommand{\STAB}[1]{\begin{tabular}{@{}c@{}}#1\end{tabular}}

\usepackage[table]{xcolor}
\usepackage{collcell}
\usepackage{hhline}
\usepackage{pgf}
\newcommand\gray{gray}

\newcommand\ColCell[1]{%
  \pgfmathparse{#1<.5?1:0}%
    \ifnum\pgfmathresult=0\relax\color{white}\fi
  \pgfmathparse{1-#1}%
  \expandafter\cellcolor\expandafter[%
    \expandafter\gray\expandafter]\expandafter{\pgfmathresult}#1}

\newcolumntype{E}{>{\collectcell\ColCell}<{\endcollectcell}}
\begin{document}
\mainmatter              
\title{Dysplasia grading of colorectal polyps through convolutional neural network analysis of whole slide images
\thanks{This project has received funding from the European Union’s Horizon 2020 research and innovation programme under grant agreement No 825111, DeepHealth Project.}}
%
\titlerunning{Dysplasia grading of colorectal polyps through CNN analysis of WSI}
\author{Daniele~Perlo\inst{1} \and
Enzo~Tartaglione\inst{1} \and
Luca~Bertero\inst{2} \and
Paola~Cassoni\inst{2} \and
Marco~Grangetto\inst{1}}
\authorrunning{Perlo~et~al.}
%
\institute{University of Torino, Computer Science dept., Torino, Italy \and
University of Torino, Pathology Unit, Dept. Medical Sciences, Torino, Italy\\
\email{\{daniele.perlo, luca.bertero\}@unito.it}}

\maketitle              

\begin{abstract}
Colorectal cancer is a leading cause of cancer death for both men and
women. For this reason, histo-pathological characterization of colorectal polyps is the major instrument for the pathologist in order to infer the actual risk for cancer and to guide further follow-up. 
Colorectal polyps diagnosis includes the evaluation of the polyp type, and more importantly, the grade of dysplasia.
This latter evaluation represents a critical step for the clinical follow-up.
The proposed deep learning-based classification pipeline is based on state-of-the-art convolutional neural network, trained using proper countermeasures to tackle WSI high resolution and very imbalanced dataset. 
The experimental results show that one can successfully classify adenomas dysplasia grade with 70\% accuracy, which is in line with the pathologists' concordance. 


\end{abstract}
\section{Introduction}
The cornerstone of conventional histo-pathological examination is the evaluation of hematoxylin \& eosin slides by trained pathologists to detect and/or quantify specific features or patterns and provide a diagnostic evaluation. Based on this premise, whole slide image (WSI) analysis approaches based on Deep Learning (DL) are well suited to address the tasks posed by the histo-pathological evaluation~\cite{Janowczyk16}. 
During the last few years, many specific challenges have been tackled: from lymph node metastasis detection~\cite{bejnordi2017diagnostic} to mitotic count~\cite{balkenhol2019deep}. The main aims of these approaches are multiple: i) improve pathologists' accuracy and thus diagnostic sensitivity; ii) speed-up the diagnostic workflow by addressing more menial, but time-consuming tasks; iii) improve diagnostic agreement by adopting standardized criteria.\\
Among the multiple fields of surgical pathology, gastrointestinal pathology is one of the most represented~\cite{gonzalez2020updates}, thus addressing this specific topic has the potential of significantly affecting the overall workflow of a pathology service. Colorectal polyps, pre-malignant lesions arising from the intestinal epithelium, are one of the most common gastrointestinal specimens submitted to histological examination. These lesions are usually collected during a colonoscopy, which represents the mainstay of colorectal cancer screening programs in many countries~\cite{bevan2018colorectal}. The development of these programs leads to a significant increase in this specific caseload of surgical pathology laboratories: the correct diagnostic assessment has far-reached consequences both for the patient and the public health systems. Indeed, a correct diagnosis is obviously important for the management of the patient, but it is now well acknowledged that different types of polyps are associated with different risks of developing metachronous invasive carcinomas during the following years~\cite{he2020long}. For this reason, specific algorithms have been established for tailoring patients' follow-up. 
Despite such clinical relevance, the concordance rates even among expert pathologists, in the diagnostic assessment of colorectal polyps, is far from optimal~\cite{denis2009diagnostic,foss2012inter,mollasharifi2020interobserver,van2011inter}.  Although the distinction between non-adenomatous and adenomatous tissue is usually reliable, the inter-observer agreement between different histological types and dysplasia grades are sub-optimal. For instance, the concordance in assessing a tubulo-villous polyp or low grade dysplasia ranged around 70\%.\\
In this work the main contributions are: i) the design of a deep learning pipeline to tackle the high dimensionality of WSI, working at single patches level; ii) the study on the physical resolutions suitable to deal automatically with the problem of classification of different colorectal polyps; iii) the study of different patch pre-processing approaches, where we find that, for the considered problem, the intensity of the dye present in the scans is the most informative feature of the tissue images.
\section{Related work}
\label{sec::relatedwork}
Only a limited number of works explored histo-pathological examination through deep learning-based analysis of digital whole slide images~\cite{korbar2017deep,song2020automatic,wei2020evaluation}. Among these works, Korbar~et~al.~\cite{DBLP:conf/cvpr/HeZRS16} present a crop-based framework, developed using a ResNet architecture to classify different types of colorectal polyps from whole-slide images. This work provides empirical suggestions the residual network architecture achieves better performance than other models. Following their previous work, Korbar~et~al. introduce a revised version of Grad-CAM (gradient driven class activation mapping)~\cite{DBLP:conf/iccv/SelvarajuCDVPB17} to visualize the attention map of the network for the annotated whole-slide~\cite{korbar2017deep}. Bychkov~et~al.~\cite{Dmitrii18} apply different architectures (convolutional and recurrent neural networks) in order to predict five-years disease survival probabilities for colorectal cancer and estimate the individual risk. 
This work explores the idea of using spatial information by feeding an LSTM network with the features extracted from image crops by a CNN. Recently, Wei~et~al.~\cite{wei2020evaluation} propose an analysis model for annotated tissue and perform a study on the generalization of neural models with external medical institutions. In such work, a hierarchical evaluation mechanism is proposed to extend the classification of tissue fragments to the entire slide.\\
These efforts show promising results, but the testing data size is small and, most importantly, they do not provide diagnosis based on both histological type and dysplasia grade.
Our aim is thus to evaluate the efficacy of a deep neural network for the automatic histo-pathological classification of colorectal polyps employing a large training cohort and assessing both polyp histological type and dysplasia grade. 
Barbano~et~al.~\cite{Barbano2021UniToPathoAL} shows how an hierarchical DL model for annotated tissue can take care of both colorectal polyps' type and relative dysplasia degree.

\section{Dataset}
\label{sec:Dataset}
WSI composing the dataset are collected within the CE project \textit{DeepHealth}~\cite{DeepHealth}.
This dataset contains all source WSI for \textit{UniToPatho}~\cite{9fsv-tm25-21} plus newer data.
Here we analyze \num{457} WSI  from colorectal cancer screening-undergoing patients.
Slide scanning is obtained through a Hamamatsu Nanozoomer S210 scanner configured at $\times$20 magnification (\SI{0.4415}{\micro\meter}/px) and stored as \texttt{.ndpi} file.
\begin{table}[tb]
    \centering
    \caption{Dataset composition.}\label{tab:slides}
    \vspace{5pt}
    \begin{tabular}{|c|c|c|c|c|c|c|c|}
        \hline
         & \ \  HP\ \ \   & NORM & TA.HG & TA.LG & TVA.HG & TVA.LG & \textbf{Total}\\
        \hline
        Slides & 62 & 30 & 34 & 232  & 44 & 55 & {\bfseries457}\\
        $R_t$ & 158 & 112 & 145 & 777  & 264 & 245 & {\bfseries1701}\\
        $A_t\left[\si{\centi\meter}^{2}\right]$ & 9.91 & 18.38 & 7.94 & 71.74  & 60.45 & 41.86 & {\bfseries 210.29}\\
        \hline
    \end{tabular}
\end{table}
Each WSI has been annotated by expert pathologists according to six classes chosen for our study: hyperplastic polyp (HP); normal tissue (NORM); tubular adenoma, high-grade dysplasia (TA.HG); tubular adenoma, low-grade dysplasia (TA.LG); tubulo-villous adenoma, high-grade dysplasia (TVA.HG) and tubulo-villous adenoma, low-grade dysplasia (TVA.LG).\\
Each slide is associated with some metadata (stored in NanoZoomer Digital Pathology Annotations~\texttt{.ndpa} file format), including a collection of Region of Interests (RoIs) associated with the corresponding class. Each RoI is determined by the pathologist and is defined by a free-hand contour, identifying the tissue area exhibiting histological findings. The number and the size of RoIs is highly variable and depends on both the tissue availability and the histological analysis. Such heterogeneity unfortunately, leads to dataset unbalancing:
the distribution of the data from $T$ tissue classes in our dataset is shown in Tab.~\ref{tab:slides}. In the table we read 
the number of WSIs, the number of ROIs $R$ and total tissue area $A_t$ for each $t$-th class, respectively. 




%

\section{Method}
\label{sec:method}
\begin{figure}[tb]
    \includegraphics[width=1.0\columnwidth]{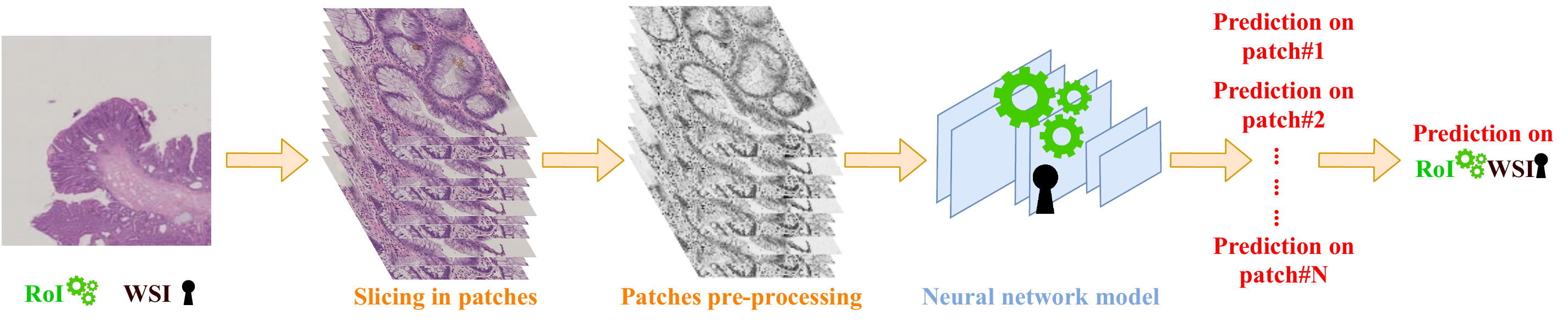}
    \caption{The neural network is trained on RoI images (gears symbol) and tested on WSI (lock symbol).}
    \label{fig:ggraph}
\end{figure}
In this section we are going to describe and motivate the proposed method. In particular, the use of deep learning for classification already proved, in similar learning tasks, to be extremely effective and robust~\cite{korbar2017deep,wei2020evaluation}. 
Direct classification on the (high resolution) whole slide, in our context, is unfeasible: the relevant features are local and can be detected at very low image scale. For this reason, the deep learning model is not trained on the full slides, but on some crops we refer to as \emph{patches}. An high-level representation of our approach is depicted in Fig.~\ref{fig:ggraph}.
Once the model is trained on patches' classification
, in order to get the whole slide classification (at validation/test time), all the scores from the single patches are averaged on the whole slide. 
WSIs have large resolution and need to be cropped into patches. 
The first operation we perform on RoIs (even before slicing them into patches) is re-scaling them to some target resolution $\varphi$.
using the Lancos-3 filter. 
Then, we slice the RoIs/WSIs into patches ($224\times224$ pixels large) using sliding windows. These patches can be immediately normalized, using approaches like \cite{macenko2009method}, or simply converting in gray-scale to reduce the expected color shift caused by hematoxylin and eosin.\\
During training we augment data: we include vertical/horizontal flips and 
a random operation chosen between rotation, equalization, solarization, inversion and contrast enhancing, as proposed in~\cite{Cubuk}.\\
In order to perform classification on the patches, we have used ResNet-18: it represents a good trade-off between complexity and performance and is one of the broadly-used to solve similar tasks~\cite{korbar2017deep,wei2020evaluation}. Pre-trained deep neural networks (on the ImageNet classification task) can be effectively used as initialization for medical classification tasks, showing good performance~\cite{korbar2017deep}.
\footnote{The pre-trained model used in all the experiments is available at \url{https://pytorch.org/docs/stable/torchvision/models.html}} 
\begin{table}[tb]
\centering
\captionsetup{justification=centering}
\caption{Dataset composition. Test RoIs are taken from a disjoint set of slides.}\label{tab:trainingtestset}
\vspace{5pt}
\begin{tabular}{|c|c|c|c|c|c|c|c|}
\hline
 & \ \  HP\ \ \   & NORM & TA.HG & TA.LG & TVA.HG & TVA.LG & Total\\
\hline
Train Slides & 50 & 25 & 26 & 203 & 36 & 45 & {\bfseries385}\\
Test Slides & 12 & 5 & 8 & 29 & 8 & 10 & {\bfseries72}\\
\hline
Train RoIs & 133 & 98 & 113 & 695 & 240 & 208 & {\bfseries1487}\\
Validation RoIs & 5 & 5 & 5 & 5 & 5 & 5 & {\bfseries30}\\
Test RoIs & 20 & 9 & 27 & 77 & 19 & 32 & {\bfseries 184}\\
\hline
\end{tabular}
\end{table}

\section{Results}
\label{sec:results}
In this section we show and discuss the classification results obtained on the WSI biopsies dataset described in Sec.~\ref{sec:Dataset} with the method proposed in Sec.~\ref{sec:method}. 
We can easily expect high error rates, considering that the information about the adenoma type is a visually global information and requires features extracted at different scale than those for the dysplasia grade, which is a more local information. Here we are not interested in distinguishing different adenoma types, but their dysplasia grade. Towards this end, we will follow a hierarchical-like classification approach~\cite{yan2015hd,zhu2017b}, 
grouping the adenoma classes into high grade (HG) and low grade (LG) dysplasia.\\
For all the experiments, we split the data at the whole slide level, in order to maintain the separation of tissues from different patients. For each class, 10\% of total patients are considered as test set. We summarise the data split in Table~\ref{tab:trainingtestset}. The validation set size is fixed to 5 RoIs for each class from the training set (likewise~\cite{wei2020evaluation}). We train our model for $250$ training epochs, and we choose the best one in terms of balanced accuracy (computed on the validation set).
Adam has been used as optimizer, and all the hyper-parameters are tuned via grid-search: 
weight decay is set to $10^{-4}$, learning rate $\eta=10^{-4}$, exponential learning rate decay $0.99$ per epoch, and minibatch size $16$. Our algorithms are implemented in Python, using PyTorch~1.5, and training/inference runs over an NVIDIA GeForce GTX~1080 GPU.\\
\subsection{Patches normalization}
\label{sec:colorstudy}
\begin{figure}[ht]
    \includegraphics[width=.5\linewidth]{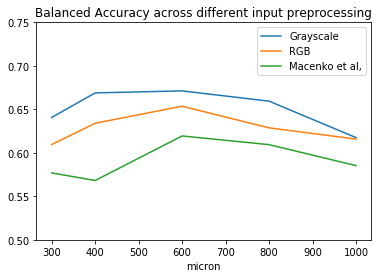}
    \centering
    \caption{Patches classification performance.}
    \label{fig:resmac}
\end{figure}
As a first step, we perform a study at different RoI resolutions: the goal here is to identify the best scale the deep model is able to extract the features. Towards this end, we consider 8 possible patches resolutions $\varphi \in [300; 1000]~\SI{}{\micro\meter}$, and 3 possible input preprocessing strategies: use of the original patches (RGB), conversion to gray-scale (gray) and the use of a standard slide normalization strategy (Macenko~et~al.~\cite{macenko2009method}), resulting in 24 training possibilities, which are reported in Fig.~\ref{fig:resmac}. For our classification task, the use of gray-scale images does not remove useful information (which might be embedded in the color) and, on the contrary, helps in removing the expected color bias~\cite{Mahapatra,ROY201842}. From our results we learn that, for the particular classification task we aim at solving, the relevant features are embed in the image texture and the signal strength, while the direct use of the RGB image does not compensate the color bias, or even standard slide normalization strategies like~\cite{macenko2009method} destroy some useful information which is not embed in the color feature. For these reasons, we will focus our analysis using gray-scale patch images as input for our model.
\subsection{Study on patches resolution for WSI classification}
\label{sec:multiscalegrey}
\begin{figure}[ht]
    \centering
    \begin{subfigure}{.45\linewidth}
      \centering
      \includegraphics[width=1.0\linewidth]{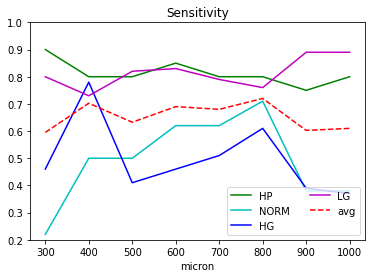}
      \caption{~}
      \label{fig:sens}
    \end{subfigure}
    \begin{subfigure}{.45\linewidth}
      \centering
      \includegraphics[width=1.0\linewidth]{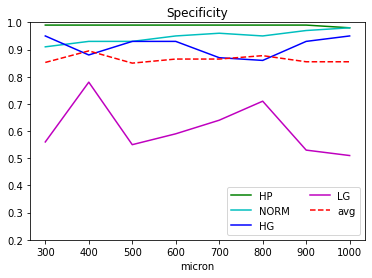}
      \caption{~}
      \label{fig:spec}
    \end{subfigure}\\
    \begin{subfigure}{.45\linewidth}
      \centering
      \includegraphics[width=1.0\linewidth]{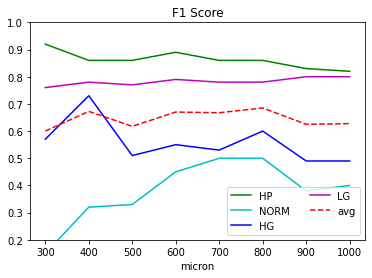}
      \caption{~}
      \label{fig:lgdysp}
    \end{subfigure}%
    \begin{subfigure}{.45\linewidth}
      \centering
      \includegraphics[width=1.0\linewidth]{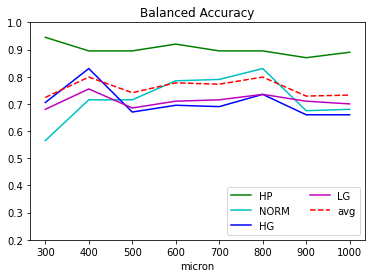}
      \caption{~}
      \label{fig:bacc}
    \end{subfigure}
    \caption{WSI inference performance comparison between different tissue categories at different patches resolutions: sensitivity (a), specificity (b), F1-score (c) and balanced accuracy (d). Red dashed line is the average performance (avg).}
    \label{fig:performances}
\end{figure}
\noindent Here we will inspect more in depth the study on WSI classification performance using gray-scaled input. Fig.~\ref{fig:performances} provides a general overview of some metrics evaluated.
There is not a clear choice regarding the optimal scale features have to be extracted. If our goal is to maximize the sensitivity for the HG class, we should choose \SI{400}{\micro\meter}: inspecting the HP's specificity for the same scale, we observe a drop which, however, is overall tolerable. F1-score gives us a more global information: indeed, for the HG class, \SI{400}{\micro\meter} is the best one. However, if we look at average performance on all classes (avg), focusing on F1-score and balanced accuracy, we can observe similar performance for \SI{400}{\micro\meter} and 600-\SI{800}{\micro\meter}.\\
\begin{table}[tb]
    \centering
    \caption{Human dysplasia diagnostic performance comparison}\label{tab:perfPath}
    \vspace{5pt}
    \begin{tabular}{|c|c|c|c|c|}
        \hline
         & & Accuracy & Sensitivity & Specificity \\
        \hline
        \multirow{3}{*}{Hyperplastic} & Our (\SI{400}{\micro\meter}) & 0.90 & 0.80 & \textbf{0.99} \\
         & Our (\SI{600}{\micro\meter}) & \textbf{0.92} & \textbf{0.85} & \textbf{0.99} \\
         & Pathologist~\cite{denis2009diagnostic} & 0.79 & 0.30 & 0.97 \\
        \hline
        \multirow{3}{*}{Low Grade} & Our (\SI{400}{\micro\meter}) & \textbf{0.76} & 0.73 & \textbf{0.78} \\
         & Our (\SI{600}{\micro\meter}) & 0.71 & \textbf{0.83}  & 0.59 \\
         & Pathologist~\cite{denis2009diagnostic} & 0.66 & 0.57 & 0.69 \\
        \hline
        \multirow{3}{*}{High Grade} & Our (\SI{400}{\micro\meter}) & \textbf{0.83} & 0.78 & 0.88\\
         & Our (\SI{600}{\micro\meter}) & 0.70 & 0.46 & \textbf{0.93} \\
         & Pathologist~\cite{denis2009diagnostic} & \textbf{0.83} & \textbf{0.81} & 0.84\\
        \hline
    \end{tabular}
    \vspace{-8pt}
\end{table}
It is important to compare the model performance with the results obtained by human pathologists. Table~\ref{tab:perfPath} reports performance comparison for HP, LG and HG in terms of balanced accuracy, sensitivity and specificity. Here, human pathologist's average performance is taken from Denis~et~al.'s work~\cite{denis2009diagnostic}, evaluated on qualitatively similar data. As we observe, our performance is very close to the pathologists'. In particular, HP classification increases of more than 10\% in accuracy, showing a quite significant improvement in terms of sensitivity. LG classification improves as well up to 10\% in balanced accuracy, yielding a significant improvement both in terms of sensitivity and specificity. HG classification score is in the same order than human pathologists (this finding is likely to be due to  HG features that are known to be visually easier to detect).

\subsection{WSI classification with 600\si{\micro\meter} patches}
\begin{figure}[ht]
    \centering
    \subcaptionbox{NORM}{
    \begin{subfigure}{.23\linewidth}
      \centering
      \includegraphics[width=0.99\linewidth]{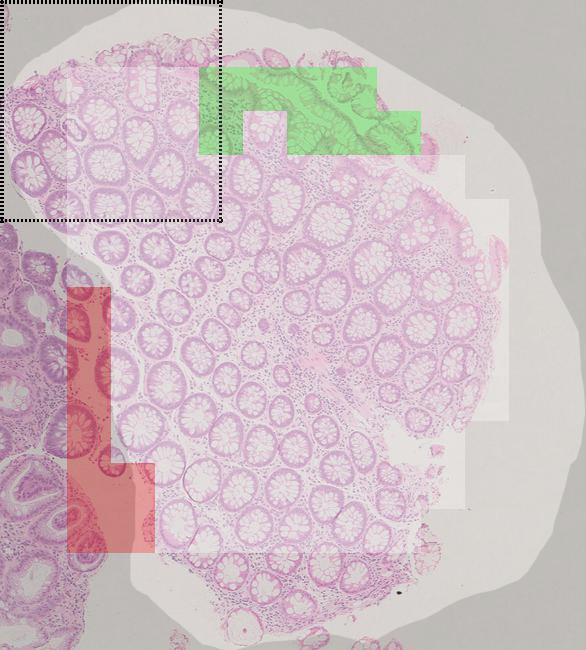}
      \label{fig:sub1}
    \end{subfigure}
    }
    \subcaptionbox{HP}{
    \begin{subfigure}{.23\linewidth}
      \centering
      \includegraphics[width=0.99\linewidth]{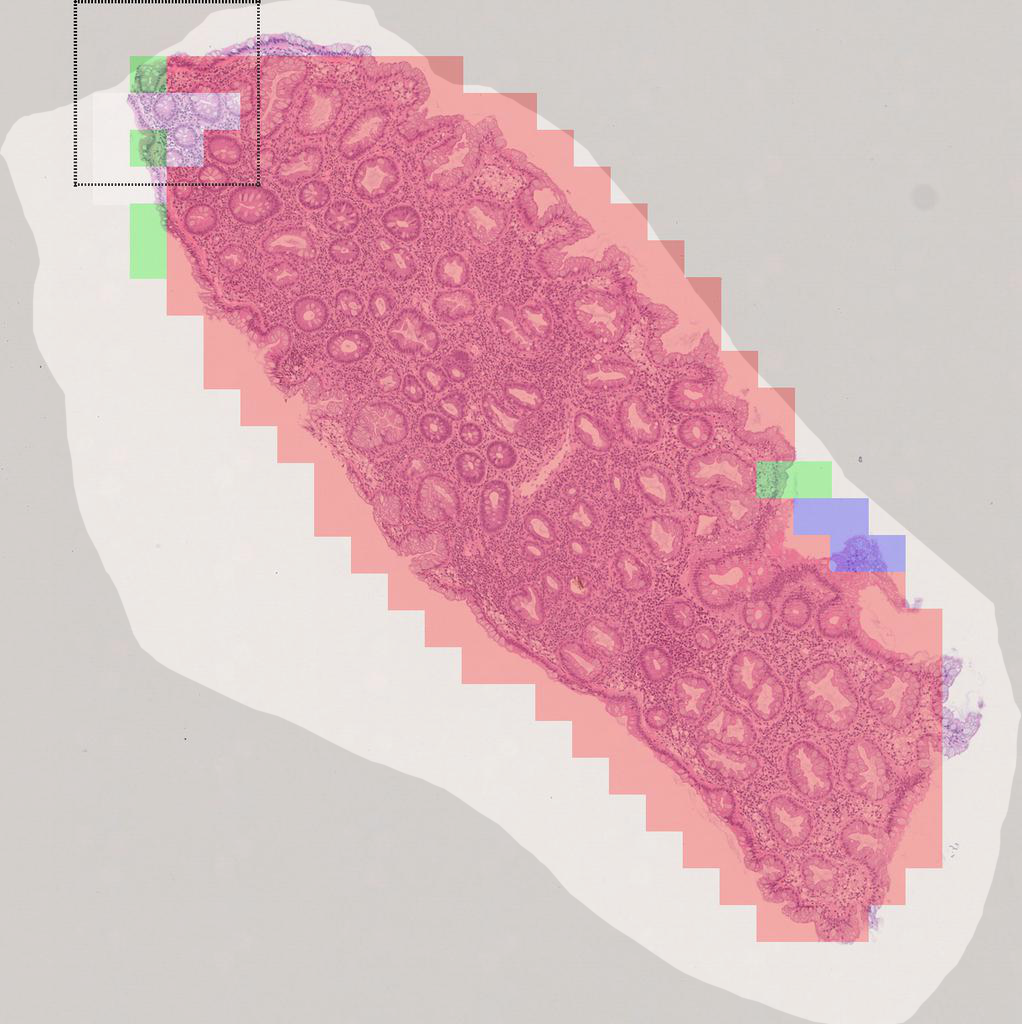}
      \label{fig:sub2}
    \end{subfigure}
    }
    \subcaptionbox{LG}{
    \begin{subfigure}{.23\linewidth}
      \centering
      \includegraphics[angle=90,origin=c,width=0.99\linewidth, trim=150 0 150 0, clip]{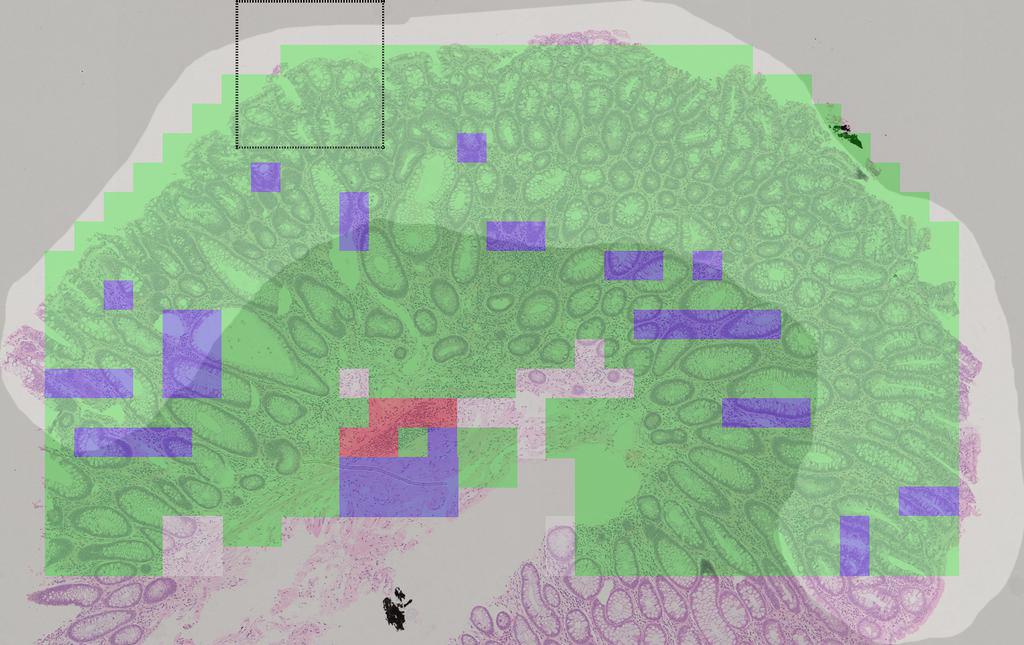}
      \label{fig:sub3}
    \end{subfigure}%
    }
    \subcaptionbox{HG}{
    \begin{subfigure}{.23\linewidth}
      \centering
      \includegraphics[width=0.99\linewidth]{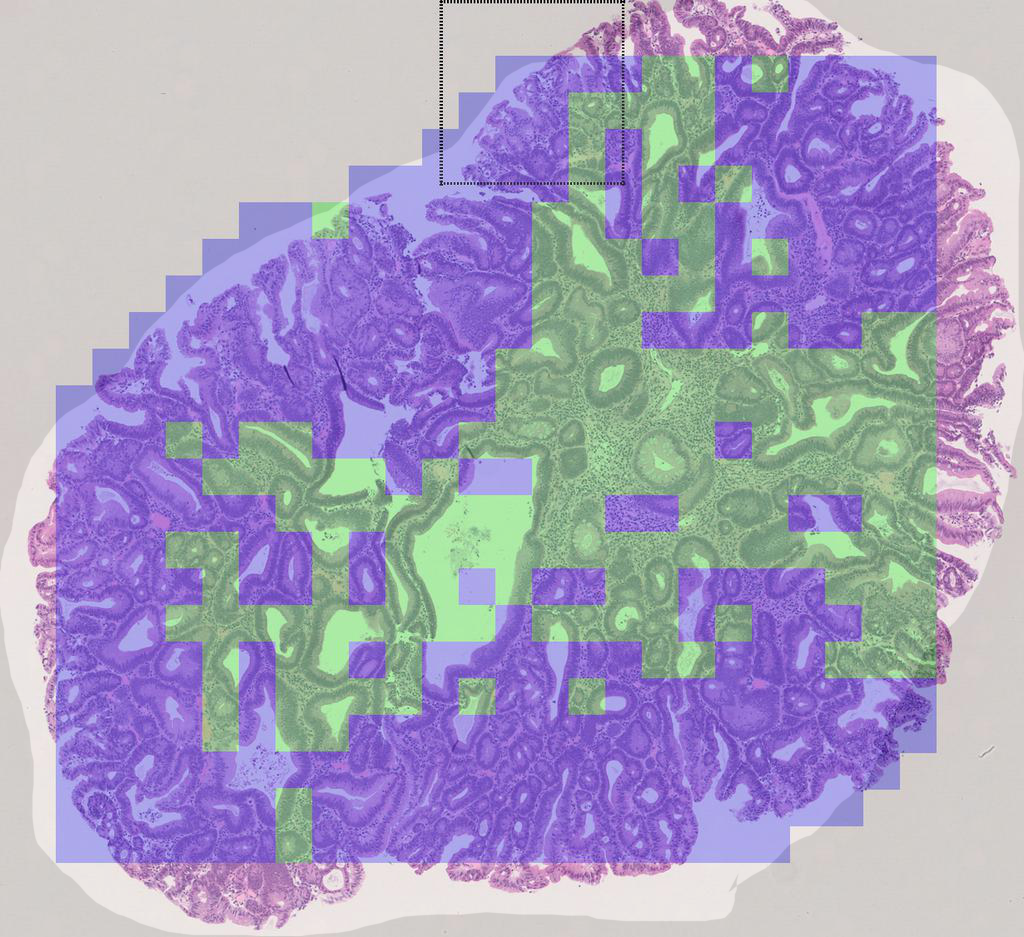}
      \label{fig:sub4}
    \end{subfigure}
    }
    \caption{Patch classification: 
    each box is located at the center of the corresponding patch with a color representing the predicted class: HP (red), NORM (white), LG (green), HG (blue). The black dashed square visually represents the patch scale ($\varphi=$\SI{600}{\micro\meter}). }
    \label{inferenceExamples}
    \vspace{-8pt}
\end{figure}
\noindent Considering that the overall performance shown by 400, 600, 700 and \SI{800}{\micro\meter} is similar, we decided here on to focus on $\varphi=\SI{600}{\micro\meter}$. Such a scale is a fair compromise, considering that other works in the literature focus on similar scales ~\cite{korbar2017deep,wei2020evaluation}. Fig.~\ref{inferenceExamples} reports a patch-level classification result for the four possible WSI classes. In particular, we observe that the model finds some HG patches within the LG WSI (Fig.4c), and viceversa (Fig.4d). This is an expected behavior, given that the dysplasia grade is provided by the pathologists according to the quantity of tissue (in our case, the number of patches) with high-grade dysplasia.\\
\begin{table}[tb]
    \caption{WSI inferences: confusion matrices.}
    \label{tab:perfInf4g}
    \begin{subtable}{.5\linewidth}
      \centering
        \caption{$\varphi=$\SI{600}{\micro\meter}, gray-scale}
\begin{tabular}{c |c|c|c|c|c|}
        \multicolumn{2}{c}{~}&\multicolumn{4}{c}{Predicted}\\
        \cline{3-6}
        \multicolumn{2}{c|}{~}&~~~HP~~~ & NORM  & ~~HG~~ & ~~LG~~ \\
        \cline{2-6}
        \raisebox{5pt}{\multirow{4}{*}{\STAB{\rotatebox[origin=c]{90}{Gr. truth~~~}}}}
        ~&HP & \textbf{0.85} & 0 & 0.05 & 0.1 \\ 
        \cline{2-6}
        ~&NORM & 0.12 & \textbf{0.75} & 0 & 0.12  \\ 
        \cline{2-6}
        ~&HG & 0.02 & 0 & \textbf{0.63} & 0.35  \\ 
        \cline{2-6}
        ~&LG & 0.03 & 0.09 & 0.18 & \textbf{0.7}   \\
        \cline{2-6}
    \end{tabular}
    \end{subtable}%
    \begin{subtable}{.5\linewidth}
      \centering
        \caption{$\varphi=$\SI{600}{\micro\meter}, RGB}
\begin{tabular}{c |c|c|c|c|c|}
        \multicolumn{2}{c}{~}&\multicolumn{4}{c}{Predicted}\\
        \cline{3-6}
        \multicolumn{2}{c|}{~}&~~~HP~~~ & NORM  & ~~HG~~ & ~~LG~~ \\

        \cline{2-6}
        \raisebox{5pt}{\multirow{4}{*}{\STAB{\rotatebox[origin=c]{90}{Gr. truth~~~}}}}
        ~&HP & \textbf{0.75} & 0.05 & 0 & 0.2 \\ 
        \cline{2-6}
        ~&NORM & 0 & \textbf{0.62} & 0 & 0.38  \\ 
        \cline{2-6}
        ~&HG & 0 & 0.02 & \textbf{0.61} & 0.37  \\ 
        \cline{2-6}
        ~&LG & 0.03 & 0.06 & 0.15 & \textbf{0.76}   \\
        \cline{2-6}
    \end{tabular}
    \end{subtable} 
\end{table}
At $\varphi=\SI{600}{\micro\meter}$, the classification between TA and TVA classes in general is poor: this is due to the larger scale required to extract proper features for adenoma classification. This, however, is not our goal, since we are here interested in classifying the dysplasia grade. Hence, we group HG and LG and we obtain the confusion matrix shown in Table~\ref{tab:perfInf4g} on WSI: the score is competitive to the human classification, as described in Sec.~\ref{sec:multiscalegrey}. We also report the confusion matrix for the equivalent model, using RGB images: as also observed in Sec.~\ref{sec:colorstudy}, the use of gray-scale images positively impacts on the WSI inference task.
\begin{figure}[tb]%
    \centering
    \subcaptionbox{NORM}{
    \begin{subfigure}{.48\linewidth}
      \centering
      \includegraphics[width=.95\linewidth]{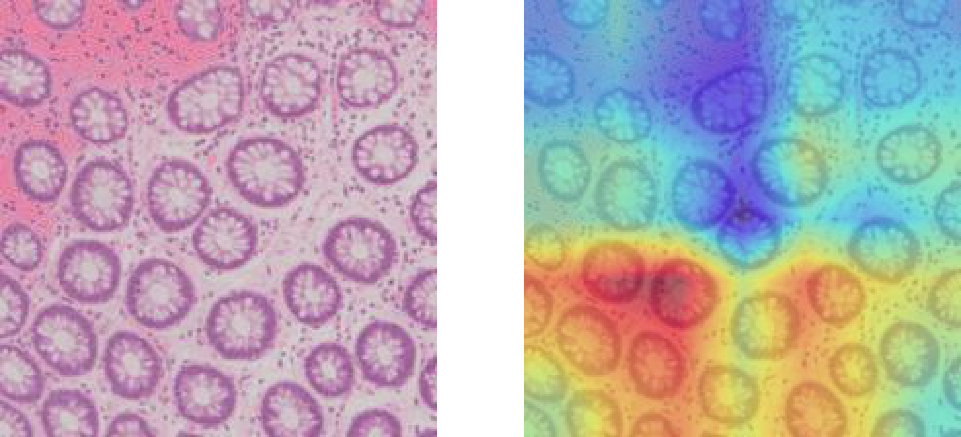}
    \end{subfigure}
    }~
    \subcaptionbox{HP}{
    \begin{subfigure}{.48\linewidth}
      \centering
      \includegraphics[width=.95\linewidth]{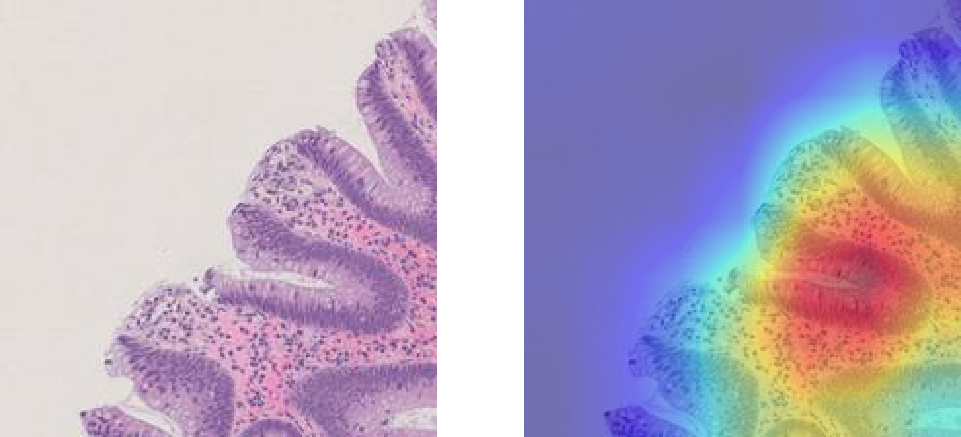}
    \end{subfigure}
    }\\
    \subcaptionbox{LG}{
    \begin{subfigure}{.48\linewidth}
      \centering
      \includegraphics[width=.95\linewidth]{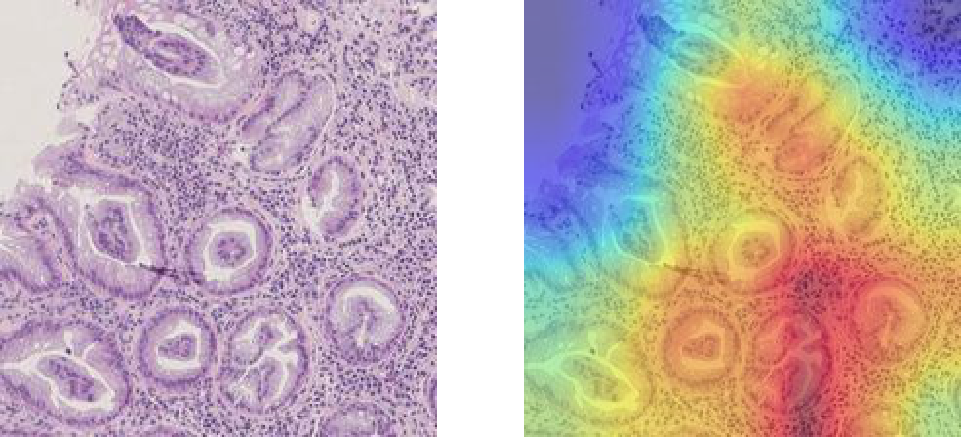}
    \end{subfigure}
    }~
    \subcaptionbox{HG}{
    \begin{subfigure}{.48\linewidth}
      \centering
      \includegraphics[width=.95\linewidth]{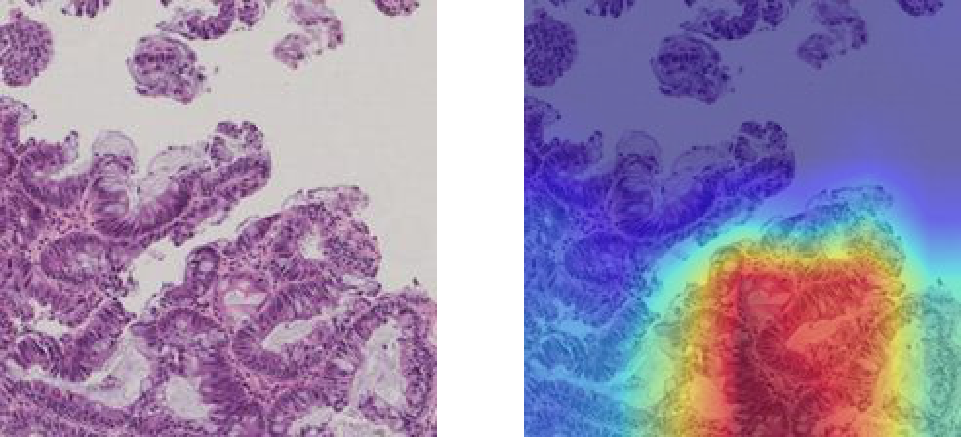}
    \end{subfigure}
    }
    \caption{Regions where the trained neural network model focuses on \SI{600}{\micro\meter} patches.}
    \label{fig:gradcam}
\end{figure}
Additionally, we inspect the areas our deep model focuses in order to perform classification by using Grad-CAM. Fig.~\ref{fig:gradcam} shows that areas of focus are consistent with the most relevant features of each histo-pathological category. For example, the hot spot of the HP sample is on a serrated gland which is a characteristic finding of this entity.

\section{Conclusion}
\label{sec:conclusion}
In this work we have designed a neural network-based pipeline for the classification of colorectal polyps in histopathological slides. We found performance benefits by applying grayscale Luma transformation~\cite{Luma} to input tissue patches.\\
We focused on four tissue classes: normal, hyperplastic, high-dysplasia and low-dysplasia adenoma. The dysplasia degree of adenomas is a very important evaluation element for the histopathologist because it leads to different post-polypectomy surveillance protocols~\cite{hassan2020post}.
The collected data enable a classification on the dysplasia degree in adenomas. The classification is performed by ResNet-18, inspecting WSI in single patches, and then classified averaging scores on all the patches. Our experiments show a performance which is very close to human pathologists~\cite{denis2009diagnostic}. Future work includes the design of a neural network model able to learn to  extract relevant tissue RoIs from the whole slide, evaluated by pathologists' agreement.
%
%
\bibliographystyle{splncs03}
\bibliography{author}
\end{document}